\documentclass[journal,twoside,web]{ieeecolor}
\usepackage{tmi}
\usepackage{cite}
\usepackage{amsmath,amssymb,amsfonts}
\usepackage{algorithmic}
\usepackage{graphicx}
\usepackage[table,xcdraw]{xcolor}
\usepackage{tabularx}
\usepackage{multirow}
\usepackage[switch]{lineno}
\usepackage{textcomp}
\usepackage{xspace}
\usepackage[T1]{fontenc}

\def\BibTeX{{\rm B\kern-.05em{\sc i\kern-.025em b}\kern-.08em
    T\kern-.1667em\lower.7ex\hbox{E}\kern-.125emX}}
\begin{document}
\title{sMRI-PatchNet: A novel explainable patch-based deep learning network for Alzheimer's disease diagnosis and discriminative atrophy localisation with Structural MRI}
\author{Xin Zhang, Liangxiu Han*,Lianghao Han,  Haoming Chen, Darren Dancey, Daoqiang Zhang
\thanks{Xin Zhang, Liangxiu Han, Darren Dancey are with the Department of Computing, and Mathematics, Manchester Metropolitan University, Manchester M15GD, U.K (e-mail: x.zhang@mmu.ac.uk; l.han@mmu.ac.uk, d.dancey@mmu.ac.uk)}
\thanks{Lianghao Han is with the Department of Computer Science,  Brunel University, Uxbridge UB8 3PH, U.K (e-mail:lianghao.han@brunel.ac.uk)}
\thanks{Haoming Chen is with the Department of Computer Science,  University of Sheffield, Sheffield S1 4DP U.K (e-mail:hchen78@sheffield.ac.uk)}
\thanks{Daoqing Zhang is with College of Computer Science and Technology, Nanjing University of Aeronautics and Astronautics, P.R.China (e-mail:dqzhang@nuaa.edu.cn)}
\thanks {Corresponding author*: L. Han (e-mail: l.han@mmu.ac.uk)}
}
\maketitle
\begin{abstract}
Structural magnetic resonance imaging (sMRI) can identify subtle brain changes due to its high contrast for soft tissues and high spatial resolution. It has been widely used in diagnosing neurological brain diseases, such as Alzheimer’s disease (AD). However, the size of 3D high-resolution data poses a significant challenge for data analysis and processing. Since only a few areas of the brain show structural changes highly associated with AD, the patch-based methods dividing the whole image data into several small regular patches have shown promising for more efficient sMRI-based image analysis. The major challenges of the patch-based methods on sMRI include identifying the discriminative patches, combining features from the discrete discriminative patches, and designing appropriate classifiers. This work proposes a novel patch-based deep learning network (sMRI-PatchNet) with explainable patch localisation and selection for AD diagnosis using sMRI. Specifically, it consists of two primary components: 1) A fast and efficient explainable patch selection mechanism for determining the most discriminative patches based on computing the SHapley Additive exPlanations (SHAP) contribution to a transfer learning model for AD diagnosis on massive medical data; and 2) A novel patch-based network for extracting deep features and AD classfication from the selected patches with position embeddings to retain position information, capable of capturing the global and local information of inter- and intra-patches. This method has been applied for the AD classification and the prediction of the transitional state moderate cognitive impairment (MCI) conversion with real datasets. The experimental evaluation shows that the proposed method can identify discriminative pathological locations effectively with a significant reduction on patch numbers used, providing better performance in terms of accuracy, computing performance, and generalizability, in contrast to the state-of-the-art methods.
\end{abstract}


\section{Introduction}
\label{sec:introduction}
\IEEEPARstart{A}lzheimer's disease (AD) is a degenerative brain illness and the most prevalent cause of dementia, accounting for 60\% to 80\% of cases \cite{20212021}. Currently, there is no cure for AD. However, an accurate and timely AD diagnosis can give patients the best chance to prepare a treatment plan that may change the disease progression and reduce the symptom\cite{Rasmussen2019Alzheimer}. So far, brain atrophy\cite{Fox1999Serial}, gray matter atrophy \cite{Lin2021Linking}, and regional atrophy \cite{Blennow2018Biomarkers}, are considered as the most critical neurodegeneration bio-markers. Structural magnetic resonance imaging (sMRI) as a non-invasive method measures brain morphometry, and is able to capture the subtle brain changes induced by the atrophic process, thanks to its high contrast for soft tissues and high spatial resolution \cite{Frisoni2010clinical}. It has been used in detecting AD and moderate cognitive impairment (MCI) with various computer vision methods\cite{Frisoni2010clinical,Lundervold2019overview,Vemuri2010Role}.  The existing sMRI-based AD diagnostic methods usually partition the entire MR image into multiple regions for better feature extraction of local abnormal brain structural changes \cite{Arbabshirani2017Single,Falahati2014Multivariate,Leandrou2018Quantitative,Rathore2017review,Zhang2021Explainable}. Depending on the partition scale, it can be broadly grouped into three categories: 1) Voxel-based; 2) Regions of interest (ROIs)-based and 3) Patch-based methods. 
\par The voxel-based sMRI diagnostic methods take whole images as input and extract global voxel-wise features for AD diagnosis\cite{Baron2001In,Khvostikov20183D,Moller2016Alzheimer}. Features such as the probability maps of gray matter (GM) \cite{Baron2001In}, white matter (WM) \cite{Kao2019White}, and cerebrospinal fluid (CSF) \cite{Ossenkoppele2015Cerebrospinal} are widely used. However, there exist some limitations \cite{Hara2018Can} including 1) Due to the high dimensionality of the feature extracted from the data, the number of data used for model training is relatively small, resulting in computationally intensive and over-fitting. 2) Ignoring area information of brain that has been shown to be important in the diagnosis of AD. 

To alleviate the aforementioned problems, several existing works focused on some predetermined ROIs guided by prior biological knowledge and extracted regional features for AD diagnosis \cite{Ahmed2015Classification,Ahmed2017Recognition,Gutman2009Disease,Magnin2009Support,Planche2017Hippocampal,Wang2007Large}. However since these methods are based on empirical regions, they might neglect possible pathological locations in the whole brain \cite{Lian2020Hierarchical}. The features extracted from the ROI may not capture the microscopic variation that is involved in the brain \cite{Zhang2016Detecting}. Additionally, segmenting ROIs based on expert knowledge is resource intensive. 

\par To locate the subtle brain atrophy regions for the early diagnosis of AD and avoid fine-grained segmentation, patch-based methods, as a compromise between voxel-based and ROI-based methods, are proposed for the effective capture of small local structural changes in sMRI images. Unlike the other two types of methods, this type of methods segments an sMRI image into multiple small fixed-size regular 3D patches based on prior-knowledge/pre-defined anatomical landmarks \cite{Gordon2014Regional,Lian2020Hierarchical,Suk2014Hierarchical} or statistics methods \cite{Zhu2021Dual}. However, the pre-defined patch selection approach may miss some AD-related atrophy patches. Statistic analysis such as the T-test does not necessarily identify the correct regions linked with AD due to carry-over effects and lacks explainability \cite{Sturdevant2021Statistical}. In these approaches, the feature extraction and classification from the selected patches are often based on conventional machine learning methods such as Support Vector Machines (SVM) \cite{Hearst1998Support}, and Linear programming boosting \cite{Demiriz2002Linear}. In \cite{Moradi2015Machine,Othman2011MRI,Zhang2018Analysis}, the authors used the handcrafted features, which may degrade the classification performance due to the heterogeneity between features and subsequent classification algorithms. Recently, deep learning methods based on convolutional neural networks (CNNs) for AD diagnosis directly learn feature representations from input patches without needing feature selection \cite{Lian2020Hierarchical, Liu2012Ensemble,Suk2014Hierarchical,Zhu2021Dual,Wang2017Comparison}, making the whole process much more convenient and less prone to error and bias.  However, in these methods, an image is firstly partitioned into patches and then each selected input patch is fed into the CNN independently. The local position information of each patch and the spatial relationship between patches are not included in the CNN computation. Few deep learning methods with the black-box nature of neural networks have specific output functions for pathological location positioning. As cerebral atrophy typically appears to be localised, only a few areas on sMRI scanning have significant structural changes highly correlated with the pathological characteristics, while the remaining areas give few useful distinguishing information. Hence, two remaining challenges in the patch-based methods are 1) how to accurately locate and select the patches; 2) how to capture both local and global features for improved explainable AD diagnosis. 
To address these challenges, this work proposes a novel patch-based neural network (sMRI-PatchNet) with explainable patch localisation and selection for Alzheimer's disease diagnosis and discriminative atrophy using Structural MRI. Our contributions include:

\begin{enumerate}
  \item An explainable patch localisation and selection for discriminative atrophy regions is proposed, in which the fewest number of AD-related patches with explainability are selected based on a novel fast recursive partition perturbation method for computing the SHapley Additive exPlanations (SHAP) contribution to a transfer learning model for AD diagnosis on massive medical data. This significantly reduces computational complexity and enhances explainability.
  \item A novel patch-based deep learning model (sMRI-PatchNet) is proposed for improving AD diagnosis performance, in which a learned position embedding is added to the patch presentation to retain the position and spatial relationship of patches. The sMRI-PatchNet has three main parts: Global spatial information (GSI) and Local patch information (LPI), are used to capture global feature between patches and local features within a patch efficiently, and a classifier for feature classification.
  \item The proposed approach has been evaluated against real datasets with the corresponding visualization. From a clinical perspective, the visualization results of brain regions covered by selected patches show that the proposed method can effectively identify discriminative pathological locations. These new biomarkers can help clinicians in clinical diagnosis.
\end{enumerate}
\par The remaining part of this paper is organized as follows: Section 2 presents the related work; Section 3 details the proposed method; Section 4 and 5 describe the experimental evaluation and results. Section 6 provides the discussion on the potential of clinical translation and the limitations of the proposed work; Section 7 concludes the work.

\section{Related work}

\par In this section, we review patch-based brain diagnosis methods using sMRI and explainable artificial intelligence methods.

\subsection{Patch-based brain diagnosis methods in sMRI}
\par Based on partitioning scales, the existing sMRI-based AD diagnostic methods can be broadly divided into three categories: 1) voxel-based, 2) regions of interest (ROIs) based, and 3) patch-based methods.  
The voxel-based methods are intuitive and straightforward in terms of the interpretation of results, aiming to identify disease-related microstructures from sMRIs of patients. The key point of this type of methods is to find suitable image features to estimate the probability of different tissue classes in a given voxel, such as gray matter (GM), white matter (WM), and cerebrospinal fluid (CSF) \cite{Kloppel2008Automatic}. However, only analysing the features of isolated voxels would lead to the ignorance of high correlations between voxels. Another limitation of voxel-level methods is the overfitting problem because the voxel-level feature representations always have higher dimensionality compared with the number of image samples in model training. Several feature dimensionality reduction algorithms are used to solve this issue, e.g. a sparse coding method with a hierarchical tree-guided regularisation \cite{Liu2012Tree-guided}. An alternative solution to feature extraction is to use 3D CNN. In \cite{Korolev2017Residual}, the authors have demonstrated that using 3D CNNs to extract features for AD classification can achieve better accuracy than traditional hand-crafted feature extraction approaches. In \cite{Zhang2021Explainable}, the authors have designed a self-attention 3D CNN to improve the diagnosis performance by adding attention to global features. However, the main limitation of 3D CNN methods is their extra-high computation costs caused by 3D convolution operations. 
In contrast, ROI-based approaches are based on the predefined regions identified from prior biological knowledge, such as the shrinkage of cerebral cortices, hippocampi and ventricles etc. \cite{Gerardin2009Multidimensional,Gutman2009Disease,Planche2017Hippocampal}. These methods require a much lower feature dimensionality than the whole voxel-based methods. However, disease-related structural/functional changes occur in multiple brain regions. The ROI-based approach may neglect disease-related features or fail to capture small and subtle changes associated with brain diseases \cite{Zhang2016Detecting}. Additionally, segmenting ROIs based on experts knowledge is resource intensive, which remains a challenging task \cite{Suk2014Hierarchical}.
\par To address these limitations, the patch-based methods have been proposed, in which brain regions are split into several small fix-sized 3D patches. Regular patches eliminate the need for region segmentation in the dataset, and each patch is a region of interest. Since brain atrophy usually occurs locally, only a few of the regions in sMRI scans have noticeable structural changes, highly associated with pathological features.  The existing works have been mainly focusing on two main challenges :1) how to select patches and combine the local patches to capture global information of the whole brain sMRI?  2) how to extract representative features and classify the patches into the right categories?  

For the first challenge, empirical knowledge-based and statistical analysis-based methods have been used for the patch selection. On one hand, the empirical knowledge-based methods select the patches in the important regions based on prior knowledge. For example, Lian et al \cite{Lian2020Hierarchical} adopted anatomical landmarks defined in the whole brain image as prior knowledge for generating selected patches. These anatomic landmarks were defined using a shape constraint regression forest model \cite{Zhang2016Detecting}. On the other hand, the statistical analysis-based selection methods use statistical algorithms to calculate the patch differences between Alzheimer’s disease (AD) and  Normal cohort (NC) patients. The patches with the highest variance are selected as the discriminative patches. In previous studies \cite{Liu2012Tree-guided,Liu2014Hierarchical,Suk2014Hierarchical,Zhu2021Dual}, a T-test was used to find the difference between AD patients and NC group data for each patch. The patches with p-values smaller than 0.05 were selected. In the study \cite{Bhatia2014Hierarchical}, the authors used the weighted correlation coefficient \cite{Wachinger2010Manifold} as the similarity measure to select discriminative patches. However, the statistical significance for voxels in each patch does not necessarily have a link with AD. Therefore, the explainable patch selection is still a challenging task. 

For the second challenge, research efforts have been made on the feature extraction and classification of patched data. Liu et al. \cite{Liu2014Hierarchical} first developed a patch-based AD diagnosis method with an independent feature extraction for each patch. The features were then integrated hierarchically at the classifier level. Inspired by Liu’s method, Suk et al. \cite{Suk2014Hierarchical} proposed a systematic method for a joint feature representation from the paired patches of sMRI images using a patch-based approach. Tong et al. \cite{Tong2014Multiple} developed a multiple instance learning (MIL) model for AD classification and MCI conversion prediction using local intensity patches as features. Zhu et al. \cite{Zhu2021Dual} proposed a dual attention multi-instance deep learning network (DA-MIDL) for the early diagnosis of AD, in which a Patch-Nets with spatial attention blocks was used for extracting discriminative features of each patch. It has been proven that these patch-based methods can efficiently deal with the problem of high dimensional features and sensitivity to slight brain structure changes.

However, in the patch-based approaches described above, each selected patch is fed into the CNN independently. The local position information of each patch and the spatial relationship between patches are not included in the CNN computation. Few deep learning methods with the black-box nature of neural networks have specific output functions for pathological location positioning. Therefore, accurately identifying the discriminative patches while capturing both local and global features for improved explainable AD diagnosis is still a remaining challenge in patch-level methods.

\subsection{Explainable methods}
\par Recently, machine learning (ML) methods, including deep learning (DL), have been enormously successful in various fields \cite{Alom2018History}. However, they are still seen as a “black box” model due to their multilayer nonlinear structure. These models have been criticized for lack of transparency, and their predicted results are not traceable \cite{Buhrmester2019Analysis}. Interpreting and explaining a “black box” model is extremely important in real applications. A reasonable interpretation of an ML model can increase the user’s trust and provide helpful information to improve the model. So far, there have been many general interpretation methods for ML/DL models, and have given birth to a new subfield, explainable artificial intelligence (XAI) \cite{Das2020Opportunities}. Based on the algorithmic approaches used, the XAI methods in medical image analysis for visual explanation can be categorized into two types: backpropagation-based and perturbation-based methods \cite{Velden2022Explainable}.
The backpropagation-based methods focus on the back-propagation of gradients through the neural network to highlight pixel attributions \cite{Selvaraju2016Grad-CAM:,Simonyan2013Deep}. The saliency map is the first interpretation method that generates a visual explanation using the back-propagation on the convolutional network \cite{Simonyan2013Deep}. The guided back-propagation method is another gradient-based XAI method to improve the saliency map by restricting the back-propagation of values less than 0 \cite{Springenberg2014Striving}. Class Activation Mapping (CAM) is also a widely used XAI method. The CAM replaces the last fully connected layers with convolutional layers to keep the object positions. This operation help discover the spatial distribution of discriminative regions for the predicted category \cite{Zhou2016Learning}. In the paper \cite{Zhang2021Explainable}, the authors used the CAM method to explain the deep learning model’s decision on AD diagnosis. However, the backpropagation-based methods are criticized for being inherently dependent on the model and data-generating process \cite{Adebayo2018Sanity}. Ghorbani et al. \cite{Ghorbani2018Interpretation} and Kindermans et al.\cite{Kindermans2019(Un)reliability} have shown that small perturbations or simple transformations to the input generated much more significant changes in the interpretations than the backpropagation-based methods did. 
The perturbation-based XAI methods focus on perturbing the input to assess the attribution of pixels in certain areas. The feature set of the input is perturbed through occlusion, removing, masking, conditional sampling, and other techniques. Then, the forward pass of the perturbed input is used to generate the attribution representations without the need for backpropagating gradients \cite{Covert2020Explaining,Ribeiro2016"Why}. The Local Interpretable Model-Agnostic Explanations (LIME) is one of the most widely used perturbation-based XAI methods because it can explain any classifier in an interpretable and faithful manner \cite{Ribeiro2016"Why}. To generate a representation that explains the model’s decision, LIME tries to find the importance of contiguous super-pixels in an input image towards the output class. Shapley additive explanations (SHAP) is a similar method that uses the classical Shapley values from game theory to show the importance to the models \cite{Shapley1951Notes,Lundberg2017Unified}. However, the perturbation-based XAI methods have the challenge of combinatorial complexity explosion. This happens when one attempts to go through all elements of the input and all their possible combinations to observe how each of them would affect the output \cite{Khakzar2019Improving}. The possible combinations of data perturbations increase dramatically when dealing with 3D images, causing a significant increase in computational costs. To avoid the combinatorial explosion, a fast perturbation method is proposed in this paper.

\section{The Proposed Method}
\par This study aims to propose a novel patch-based convolutional network (sMRI-PatchNet) with an explainable patch selection for AD diagnosis with sMRI images. It involves two-level classifications: Alzheimer’s disease (AD) vs. Normal cohort (NC), and progressive MCI (pMCI) vs. Stable MCI (sMCI).  

\par
\begin{figure}[h]
    \centering
    \includegraphics[width=0.45\textwidth]{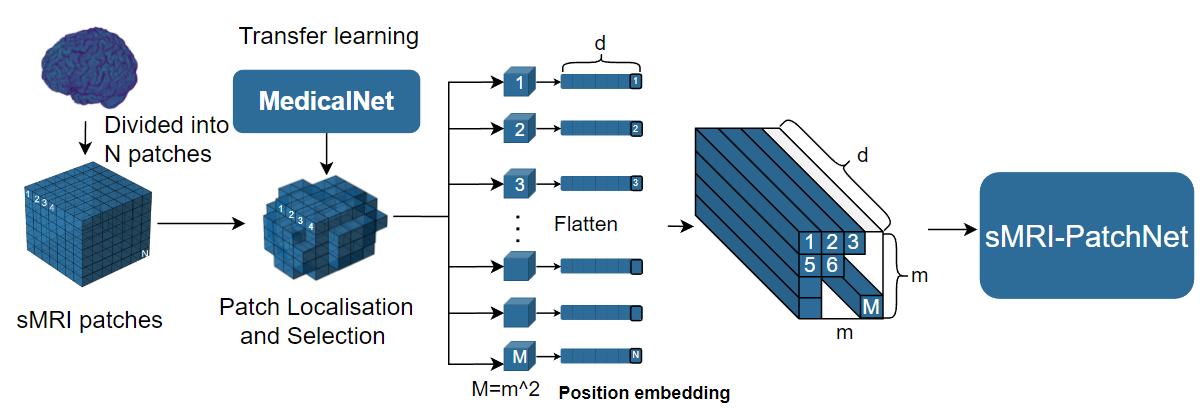}
    \caption{The flowchart of the proposed method}
    \label{FIG:1}
\end{figure}

 The schematic diagram of our framework is shown in {Fig.~\ref{FIG:1}}, which consists of two major units: Explainable Patch Localisation and Selection (EPLS) for patch selection and sMRI-PatchNet for feature extraction and classification. 
The rationale behind this architecture includes:

\begin{enumerate}
  \item Unlike traditional statistical or prior knowledge-based methods, we have proposed an explainable patch localisation and selection method. It can accurately identify the fewest number of AD-related patches based on a novel fast recursive partition perturbation method for computing the SHapley Additive exPlanations (SHAP) contribution to a transfer learning model for AD diagnosis on massive medical data. This significantly reduces computational complexity and enhances explainability.
  \item A novel patch-based deep learning network (sMRI-PatchNet) is proposed for feature extraction and classification. The selected patches are flattened into vectors using a linear projection. A learned position embedding is added to the patch presentation to retain their location and spatial information. Two CNN blocks including global spatial information (GSI) and local patch information (LPI) are proposed to capture the global features between patches and local features within a patch. A classifier consisting of average pooling followed by a fully connected layer is connected to predict output classes.
\end{enumerate}

\subsection{Explainable Patch Localisation and Selection (EPLS)}

\begin{figure}[h]
    \centering
    \includegraphics[width=0.45\textwidth]{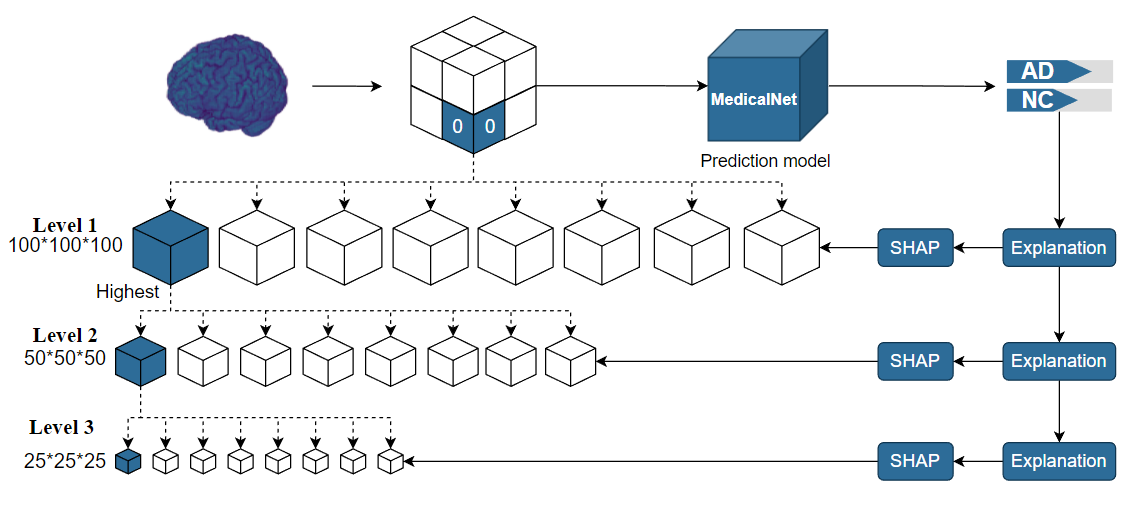}
    \caption{The flowchart of the fast recursive partition XAI method}
    \label{FIG:2}
\end{figure}

The unit of explainable patch localisation and selection (EPLS) aims to identify and select the most discriminative patches by evaluating their importance to the AD diagnosis (classification). It was implemented through transfer learning using MedicalNet \cite{Chen2019Med3D:} pretrained on 23 publicly available large medical image data, and fine-tuned with an sMRI dataset. The classification accuracy of MedicalNet can reach 0.909. 
As shown in {Fig.~\ref{FIG:2}}, in this unit, each sMRI image is uniformly partitioned into 3D cubic patches with a fixed size, without overlapping. Based on our experiments and the previous work \cite{Lian2020Hierarchical}, the size of $25 \times 25 \times 25$ is selected in this study. These patches are fed into the MedicalNet for evaluating their importance to AD diagnosis. A fast explainable recursive partition perturbation approach for assessing the patch importance based on the value of the SHAP coefficient has been designed. To explain the model’s decision, we perturb the sMRI data by filling value of 0 on specific patch and observing how the model output changes to the perturbations. Then, the SHAP coefficient is calculated to measure the contribution of each location of the sMRI input to the model output. This is aligned with human intuition and can effectually discriminate among model output classes \cite{Lundberg2017Unified}. We average the contribution of each location for all sMRI images identified as AD, and the high contribution locations are selected as input to the AD diagnosis model. 

\subsubsection{Fast recursive partition perturbation method}

\par In this work, each sMRI image is divided into n 3D cubic patches (n=598 when the patch size is $25 \times 25 \times 25$ on a standard sMRI image). The computational complexity of the model explanation on each image will be $2^n$ by using the conventional permutation method to perturb the data. This will take an unacceptable amount of time to work through all the sMRI data. 

\begin{figure}[h]
    \centering
    \includegraphics[width=0.45\textwidth]{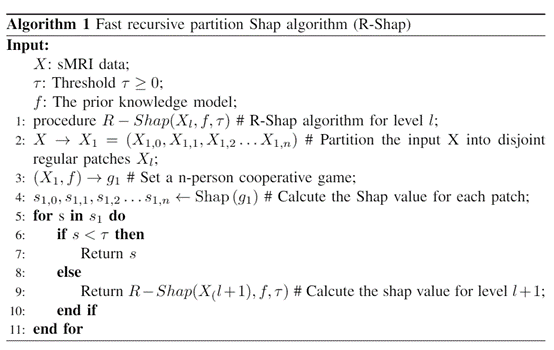}
    \caption{3.1.1.	Fast   recursive partition perturbation method}
    \label{FIG:3}
\end{figure}

To reduce the computational complexity and costs without compromising the performance, this work introduces a fast recursive partition perturbation method to perturb the image hierarchically. Unlike the conventional permutation methods which ignore the link between patches, our proposed method calculates the importance of each patch by iteratively partitioning the data. The algorithm is shown in {Fig.~\ref{FIG:3}}. Specifically, to avoid predefining regions of interest, we partition an image, $X$, into disjointed regular patches $\left(X_{1,0}, X_{1,1}, X_{1,2} \ldots X_{1, n}\right)$, called level 1, the size of each patch is 100 × 100 × 100, and n is the number of patches (n = 8 is chosen at Level 1 for 3D images). After computing the SHAP coefficients of all patches at Level 1, $\left(S_{1,0}, S_{1,1}, S_{1,2} \ldots S_{1, n}\right)$, we further partition each patch into 8 smaller sub-patches in a hierarchical manner up to Level 3, based on its SHAP value. If the calculated SHAP coefficient for a patch is greater than a pre-defined threshold ($\tau$), then there is no further partition to the next level; or else, if $S_{1, i}<\tau$, we partition this patch to the next level; and recursive this manner up to Level 3. The patch sizes are $50 \times 50 \times 50$  at Level 2 and $25 \times 25 \times 25$ at Level 3. After the recursion, all the patches along with their SHAP coefficient values are returned. 

This partitioning method matches the data structure of the image and significantly reduces the computational complexity.

\subsubsection{SHAP coefficient estimation method}
The SHAP coefficient \cite{Shapley1951Notes} is used to calculate the contribution of each patch towards the output of AD classification) and can be defined as:
\begin{equation}
S_i(f)=\sum_{C \subseteq[n] \backslash\{i\}} \frac{|C| !(n-|C|-1) !}{n !}\left[f\left(X_{C \cup\{i\}}\right)-f\left(X_C\right)\right]
\end{equation}

Where C is a subset of n patches used in the model, X is the vector of all patch features and n is the number of patches in the input. $f\left(X_{C\cup\{i\}}\right)$ is the predicted probability of AD with the $ith$ patch included while $f\left(X_C\right)$ is the probability value without the $ith$ patch in the input. $S_i(f)$ represents the averaged marginalized contribution of ith patch over all possible subsets of n. The computation complexity is exponential in the dimension of the input features ($O(2^n)$). 

\subsection{sMRI-PatchNet}

\begin{figure*}[h]
    \centering
    \includegraphics[width=0.9\textwidth]{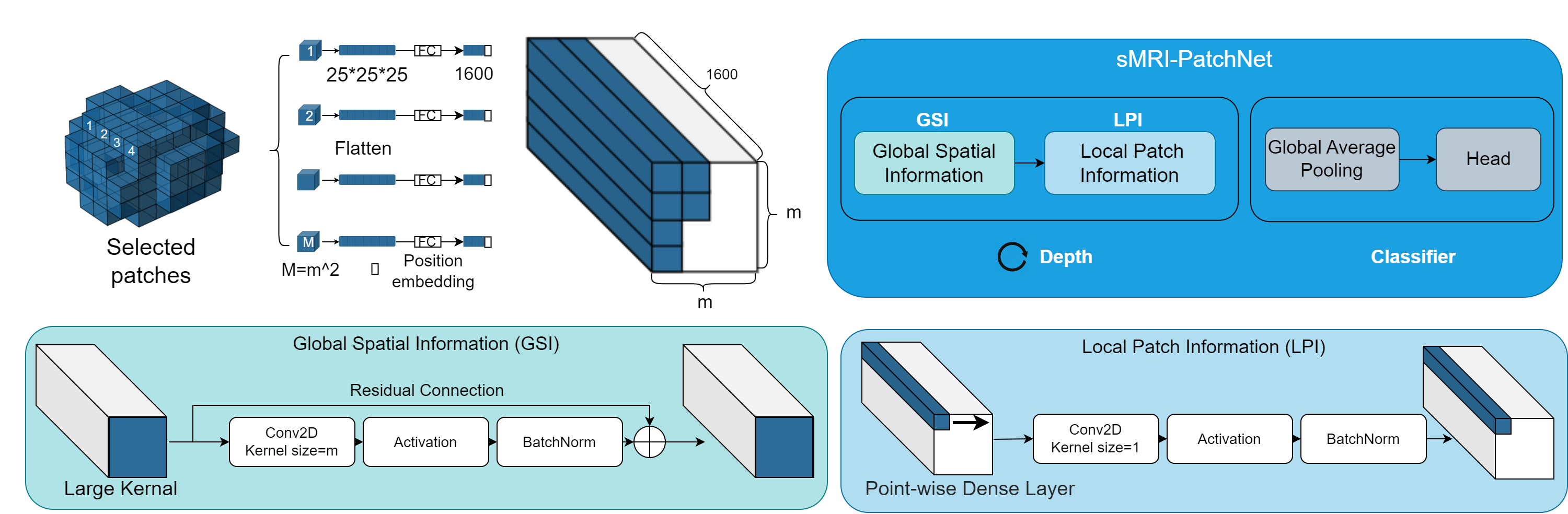}
    \caption{The architecture of the sMRI-PatchNet Model}
    \label{FIG:4}
\end{figure*}

\par The second major unit is a patch-based deep learning convolutional network for feature extraction and classification, named sMRI-PatchNet, as shown in {Fig.~\ref{FIG:4}}. After the patch selection, these selected patches are flattened into vectors ($\mathrm{x}_p \in \mathrm{R}^{M \times\left(P^3\right)}$) and mapped to d dimensions (the size of vectors) using a linear projection. Where P is the size of the patch, and M is the number of selected patches. To retain the positional information, a learned 1D position embedding \cite{Dosovitskiy2020Image} is added to the patch vectors.

\begin{multline}
\mathrm{X}=\left[\mathrm{x}_{\mathrm{p}}^1 \mathrm{E} ; \mathrm{x}_{\mathrm{p}} \mathrm{E} ; \cdots ; \mathrm{x}_{\mathrm{M}}^{\mathrm{M}} \mathrm{E}\right]+\mathrm{E}_{\mathrm{pos}}, \\
\mathrm{E} \in \mathrm{R}^{\left(\mathrm{P}^3\right) \times \mathrm{d}}, \mathrm{E}_{\mathrm{pos}} \in \mathrm{R}^{\left(\mathrm{P}^3\right) \times \mathrm{d}}
\end{multline}

We then group patch vectors in sequence to a new array ($\mathrm{X} \in \mathrm{R}^{(d+1) \times(\mathrm{m} \times \mathrm{m})}$), where $m \times m=M$ is the number of patches, d is the size of the array's Z-axis which denotes the dimension of each flattened patch. The Z-axis (d) represents the information inside each patch. The xy-axis ($m\times m$) represents the spatial information between the different small patches. The sMRI-PatchNet consists of global spatial information (GSI), local patch information (LPI), and a classifier. 
GSI is used to capture global information between the patches from the XY-axis and the LPI is used to capture the local features within a patch from Z-axis (d). The classifier consists of average pooling and a fully connected layer that classifies features into the correct class.

\subsubsection{Global spatial information (GSI)}
The GSI is proposed to capture the global spatial information of patches. In the first part, the 3D patches are flattened into vectors and then converted to a 2D array ($\mathrm{X} \in \mathrm{R}^{d \times(\mathrm{m} \times \mathrm{m})}$), which can reduce the computational consumption caused by 3D convolution. Then a spatial-wise 2D convolution with a large kernel size of $m \times m$ is used to operate on $\mathrm{X} \in \mathrm{R}^{d \times(\mathrm{m} \times \mathrm{m})}$ to extract the global spatial information. The kernel size $m\times m$ allows the receptive field of convolution to cover the entire area. Then, the conventional activation ($\sigma$, Rectified Linear Unit(ReLU) \cite{Hinton2010Rectified} ) and BatchNorm (BN \cite{Ioffe2015Batch}) are used following the GSI module to accelerate the model training. A residual connection \cite{He2015Deep} is introduced before and after the GSI to avoid the gradient vanishing problem when the depth of the model is increased.

\begin{equation}
\left.X_l^{\prime}=\mathrm{BN}\left(\text { SpatialwiseConv }\left(X_{l-1}\right)\right)\right)+X_{l-1}
\end{equation}

\subsubsection{	Local patch information (LPI)}
The LPI is proposed to capture the internal information of each patch. Note that the voxel values of each patch are flattened on the Z-axis in the input array with dimension d. Therefore, a pointwise 2D convolution with kernel size $1\times1$, regarded as a multilayer perceptron (MLP), is executed on the channel axes. The activation and Batchnorm are also followed after the LPI module.

\begin{equation}
 Z_{l+1}=\mathrm{BN}\left(\sigma\left(\text { PointwiseConv }\left(X_l^{\prime}\right)\right)\right)   
\end{equation}

\section{Experimental Evaluation}

\subsection{Dataset description and preprocessing }

This study used the ADNI dataset, which was obtained from the public Alzheimer's Disease Neuroimaging Initiative (ADNI) database (http://adni.loni.usc.edu). In the ADNI dataset, a total of 1193 1.5T/3T T1-weighted structural magnetic resonance imaging (sMRI) scans are taken from the baseline/screening visits (i.e. first examination) of subjects in the three ADNI phases (ADNI-1, ADNI-2 and ADNI-3). These participants can be classified into three groups: AD (Alzheimer's disease), MCI (mild cognitive impairment) and NC (normal controls) according to standard clinical criterias. For the prediction of MCI conversion, MCI subjects are further categorised into two groups: pMCI (progressive MCI subjects who convert to AD within 36 months of the baseline visit) and sMCI (stable MCI subjects who are consistently diagnosed with MCI). The study's ADNI dataset contained 389 AD, 172 pMCI, 232 sMCI and 400 NC subjects. The demographic details of this dataset are described in {Table.~\ref{table:1}}.

\begin{table}[h]
\caption{Demographic details of the studied subjects including dataset, group type, gender, age, mini-mental state examination (mmse) and clinical dementia rating (cdr).}\label{table:1}
\centering
\resizebox{0.45\textwidth}{!}{%
\begin{tabular}{llllll}
\hline
\multirow{2}{*}{Dataset}  & \multirow{2}{*}{Group} & Gender            & Age          & Edu          & MMSE         \\
                          &                        & (Male/Female)     & (Mean±Std)   & (Mean±Std)   & (Mean±Std)   \\ \hline
\multirow{4}{*}{ADNI-1}   & AD                     & 200 ( 103 / 97 )  & 75.62 ± 7.70 & 14.68 ± 3.20 & 23.29 ± 2.04 \\
                          & pMCI                   & 172 ( 106 / 66 )  & 76.34 ± 7.15 & 15.76 ± 2.84 & 26.61 ± 1.70 \\
                          & sMCI                   & 232 ( 154 / 78 )  & 76.47 ± 7.82 & 15.58 ± 3.17 & 27.31 ± 1.79 \\
                          & NC                     & 231 ( 119 / 112 ) & 75.99 ± 5.00 & 16.06 ± 2.84 & 29.12 ± 0.99 \\ \hline
\multirow{2}{*}{ADNI-2/3} & AD                     & 153 ( 85 / 68 )   & 74.95 ± 7.80 & 15.88 ± 2.66 & 23.03 ± 2.14 \\
                          & NC                     & 419 ( 170 / 249 ) & 74.84 ± 6.60 & 16.63 ± 2.48 & 29.09 ± 1.19 \\ \hline
\end{tabular}
}
\end{table}

The original structural MRI data from the ADNI database are pre-processed for subsequent feature learning and classification. As the original dataset is in Neuroimaging Informatics Technology Initiative (NIfTI) format, the preprocess is needed for spatial distortion correction caused by gradient nonlinearity and B1 field inhomogeneity. This is a standard pipeline process including anterior commissure (AC)-posterior commissure (PC) correction, intensity correction \cite{Sled1998nonparametric}, and skull stripping \cite{Wang2011Robust}. We have used MIPAV (Medical Image Processing, Analysis, and Visualisation) application to implement AC-PC correction and use FSL (FMRIB Software Library v6.0) for skull stripping. A linear registration strategy (flirt instruction in FSL) is also executed to align every sMRI linearly with the Colin27 template \cite{Holmes1998Enhancement} to delete global linear differences (including global translation, scale, and rotation differences), and also to re-sample all sMRIs to have the identical spatial resolution. After the preprocessing, all sMRI images have the same size, containing $181 \times 217 \times 181$ voxels.

\subsection{Evaluation metrics}
We have evaluated two binary classification tasks: AD classification (i.e., AD vs. NC) and MCI conversion prediction (i.e., pMCI vs. sMCI). The classification performance is evaluated based on four commonly used standard metrics, including classification accuracy (ACC), sensitivity (SEN), specificity (SPE), and Area under the curve (AUC). These metrics are defined as:

\begin{equation}
A C C=\frac{\mathrm{TP}+\mathrm{TN}}{\mathrm{TP}+\mathrm{TN}+\mathrm{FP}+\mathrm{FN}}
\end{equation}
\begin{equation}
S E N=\frac{T P}{T P+F N}
\end{equation}

\begin{equation}
S P E=\frac{T N}{T N+F P}
\end{equation}

Where $TP = True Positive$, $TN= True Negative$, $FP= False Positive$ and $FN=False Negative$. The AUC is calculated based on all possible SEN and 1-SPE obtained by changing the thresholds performed on the classification scores yielded by the trained networks.

\subsection{Experimental evaluation}

To evaluate the performance of our proposed model, we have conducted three types of experiments: 1) Diagnostic performance evaluation. 2) Generalisability evaluation and 3) Impact of discriminative patch location selection on model performance.

\subsubsection{Experiment One: Diagnostic performance evaluation }

In this experiment, we evaluate the diagnostic performance of our proposed model. The detailed configuration of the proposed PatchNet is shown in {Table.~\ref{table:2}}. The patch size is $25 \times 25 \times 25$, and top 36 patches with the highest SHAP value are selected. These configurations are the best combinations obtained in our experiments. The model has 16 layers (D). The dimension of the flattened patch is 1600. 

\begin{table}[h]
\caption{The model configuration}\label{table:2}
\centering
\resizebox{0.35\textwidth}{!}{%
\tiny
\begin{tabular}{ll}
\hline
Model        & sMRI-PatchNet \\ \hline
Patch size   & 25            \\
Patch number & 36            \\
Heads        & Na            \\
Depth (D)    & 12            \\
Dimension    & 1600          \\
Param (M)    & 34.53         \\ \hline
\end{tabular}
}
\end{table}

The proposed model is compared with several commonly used automatic AD diagnosis methods, including
\par 1) Three traditional machine learning (ML) based methods representing a typical example of the three types of existing computer-aided diagnostic methods for AD, respectively including:
\par a) A voxel-based method (VBM) from Ashburner et al \cite{Ashburner2000Voxel-Based}. In VBM, each sMRI is processed by the spatial normalization to a standard stereotactic brain space (i.e., Colin27 template) and the local gray matter density is measured as the voxel-level feature.
\par b) A region-based method (RBM) by Zhang et al, \cite{Zhang2011Multimodal}. The RBM uses the prior knowledge identified regions of the sMRI image as the input. After a deformable registration \cite{Shen2002HAMMER:}, an entire brain sMRI image is segmented into 93 areas according to the template with 93 manually labeled ROIs \cite{Kabani19983D}, as described in \cite{Zhang2011Multimodal}. The grey matter volume in each ROI is then calculated as a region-level feature, which is normalized by the total intracranial volume; 
\par c) A patch-based method (PBM) from Zhang et al. \cite{Liu2012Tree-guided}. The PBM uses selected patches as the input. The patch location selection method proposed in their study is used to evaluate the contribution of each patch to AD. The top 40 patches are selected, and a patch pool is operated on selected patches from tissue density maps to generate input vectors. The Light Gradient Boosting Machine (LightGBM) machine learning classifier is selected for feature classification. It has state-of-the-art accuracy, lower memory usage, and higher efficiency that can handle large-scale data \cite{Ke2017Lightgbm:}. The detailed parameters for the LightGBM classifier are shown in {Table.~\ref{table:3}}.
\par 2) One deep learning model based on transfer learning: The MedicalNet (Med3D) \cite{Chen2019Med3D:}. It is also the trained model that we used for the explainable discriminative location selection. The Med3D adopts the ResNet family (ResNet 10, ResNet 18) architecture as the backbone \cite{He2015Deep}. To enable the Med3D to train with 3D medical data, all 2D convolution kernels are replaced with their 3D versions. To avoid overfitting when trained on the limited volume of training data, Med3D collected the dataset from several medical challenges to build a large dataset and provided a pre-trained model for other downstream tasks. In this work, we use this pre-trained model and fine-tune it with sMRI data for our AD diagnosis tasks.
\par 3)  Two typical patch-based deep learning methods, HFCN \cite{Lian2020Hierarchical} and DA-MIDL \cite{Zhu2021Dual}. Both of these methods used the statistical method for patch selection and proposed novel CNN models for patch feature extraction and classification. The HFCN model is implemented by multi-layer convolutional structures. It contains three-level networks consisting of patch-level, region-level, and subject-level sub-networks. Multi-scale feature representations are jointly learned and fused for the construction of hierarchical classifiers. The features from different levels are spatially combined to feed into the classifier. The DA-MIDL model consists of three primary components: a) Patch-Nets with spatial attention blocks for extracting features within each patch; b) an attention multi-instance learning (MIL) pooling operation for balancing the relative contribution of each patch, and c) an attention-aware global classifier for further learning the features and making the AD-related classification decisions. 

\begin{table}[h]
\caption{The LightGBM classifier training parameters}\label{table:3}
\centering
\LARGE
\resizebox{0.45\textwidth}{!}{%
\begin{tabular}{lll}
\hline
Parameters      & Description                                              & Value                                    \\ \hline
Boosting\_type  & Method of boosting                                       & Gradient-boosted   decision trees (gbdt) \\
Num\_leaves     & Max number of leaves in one tree                         & 31                                       \\
Max\_depth      & Limit the max depth Forr tree model, -1 means no   limit & -1                                       \\
Num\_iterations & Number of boosting iterations                            & 100                                      \\
Learning\_rate  & The shrinkage rate for model train                       & 0.05                                     \\ \hline
\end{tabular}
}
\end{table}

In this test, we train our model using the ADNI-1 dataset and perform 10 times of five-fold cross-validation. The dataset is randomly split into five groups where four groups (80\% of the dataset) are used for training, and the rest are used for testing. The experimental results for classification performance are the average of the accuracies and and its standard deviation on the testing set across all folds. This allowed  a more appropriate model analysis and made it possible to avoid overfitting problems. To optimize model parameters, Adam \cite{Kingma2014Adam:}, a stochastic optimization algorithm with a batch size of 8 samples, has been used in training the proposed network. We first set the initial learning rate (LR) as $1\times 10^{-4}$. The LR is decreased to $1\times 10^{-6}$ with increased iterations. CrossEntropy has been selected as the loss function for this task \cite{Boer2005Tutorial}.  
The experiments are implemented based on PyTorch and executed on a server with an Intel(R) Xeon(R) CPU E5-2650, NVIDIA 2080TI, and 64 GB memory.

\subsubsection{Experiment Two: Generalisability Evaluation }
In this experiment, the generalisability and repeatability of the proposed PatchNet model are evaluated. We train the model based on the whole ADNI-1 dataset and test it on two independent datasets (ADNI-2 and ADNI-3). Due to a lack of pMCI and sMCI samples from ADNI-2 and ADNI-3, we only evaluate the model performance on the AD vs. NC classification task. The four automatic diagnosis methods described in the previous section are used for comparison.

\subsubsection{Experiment Three: The impact of discriminative patch selection on model performance}

In this experiment, we evaluate the influence of patch selections on the classification performance of our proposed classification model, based on two different patch selection approaches, our SHAP-based method, and the traditional statistic method. We investigate the effect of the patch number on classification performance and the performance of two different patch selection methods on identifying the locations of patches in the brain.  
Considering that the patch choice is based on its contribution to AD, we use the AD vs. NC classification task for evaluation. In our SHAP-based patch selection approach, the top 16, 36, and 64 patches with the highest SHAP value are selected, respectively. The number of selected patches has to be squareable to match the PatchNet requirement (that is, it can be converted to an m×m array). 
The traditional statistical analysis-based patch selection method used in \cite{Liu2012Tree-guided,Liu2014Hierarchical,Suk2014Hierarchical,Zhu2021Dual} is selected for comparison. This method assumes that the patch locations with the most significant differences between the AD and NC groups are more likely to be the brain regions with abnormal atrophy. Thus, the t-test \cite{Kim2015T} is applied to two groups of patch-level features at one patch location from the same amount of AD patients and normal controls in the training set, respectively. The p-value for each patch location is used to sort the informativeness in all patches. The patches with lower p-values are selected. Here we orderly select the number from {20, 40, 60, 120, 200} in the PatchNet and record the corresponding results.
Identifying morphological changes in the brain can help the clinical diagnosis of AD \cite{Blennow2018Biomarkers,Craig-Schapiro2009Biomarkers,Humpel2011Identifying}. To investigate the performance of the two patch selection methods on identifying the brain regions associated with AD, we quantitatively visualize the patch locations predicted as AD and the regions where they are located in the brain. Accurately identifying these regions can provide valuable information for clinical diagnosis. 

\section{Results}
\subsection{Results of Diagnostic Performance Evaluation}

\begin{table*}[h]
\caption{Results for AD classification (i.e., AD vs. NC) and MCI conversion prediction (i.e., pMCI vs. sMCI)}\label{table:4}
\centering
\resizebox{0.9\textwidth}{!}{%
\begin{tabular}{lllllllll}
\hline
\multirow{2}{*}{Model}              & \multicolumn{4}{l}{AD vs. NC classification}          & \multicolumn{4}{l}{pMCI vs. sMCI classification}      \\
                                    & ACC         & SEN         & SPE         & AUC         & ACC         & SEN         & SPE         & AUC         \\ \hline
VBM \cite{Ashburner2000Voxel-Based} & 0.815±0.043 & 0.755±0.05  & 0.873±0.036 & 0.884±0.037 & 0.682±0.054 & 0.629±0.08  & 0.714±0.06  & 0.706±0.053 \\
RBM \cite{Zhang2011Multimodal}        & 0.808±0.107 & 0.717±0.103 & 0.883±0.106 & 0.849±0.072 & 0.669±0.084 & 0.573±0.094 & 0.741±0.073 & 0.696±0.034 \\
PBM \cite{Liu2012Tree-guided} & 0.838±0.089 & 0.726±0.124 & 0.871±0.05  & 0.847±0.028 & 0.682±0.071 & 0.4±0.096   & 0.73±0.071  & 0.637±0.048 \\
Med3D-18 \cite{Chen2019Med3D:}     & 0.909±0.149 & 0.896±0.142 & 0.924±0.151 & 0.952±0.085 & 0.806±0.047 & 0.773±0.054 & 0.833±0.055 & 0.817±0.033 \\
HFCN \cite{Lian2020Hierarchical}       & 0.882±0.047 & 0.89±0.054  & 0.883±0.055 & 0.929±0.033 & 0.807±0.046 & 0.806±0.049 & 0.798±0.036 & 0.794±0.036 \\
DA-MIDL \cite{Zhu2021Dual}        & 0.904±0.079 & 0.887±0.092 & 0.903±0.075 & 0.922±0.034 & 0.809±0.092 & 0.771±0.101 & 0.826±0.097 & 0.851±0.047 \\
Our Method                          & 0.920±0.088 & 0.920±0.119 & 0.919±0.052 & 0.967±0.023 & 0.819±0.044 & 0.818±0.055 & 0.816±0.056 & 0.857±0.029 \\ \hline
\end{tabular}
}
\end{table*}

The results of AD vs. NC classification and MCI conversion prediction achieved by our sMRI-PatchNet model and the competing methods on the ADNI-1 dataset are shown in {Table.~\ref{table:4}}. The proposed sMRI-PatchNet method achieves the best accuracy (0.920 and 0.819) in the two classification tasks, which are statistically significant. Of the three machine learning-based baseline methods, the PBM outperforms the RBM and VBM methods, indicating that patch-level feature representations could offer better discriminative information regarding the subtle brain changes for brain disease diagnosis.

Moreover, as shown in {Table.~\ref{table:4}}, the Med3D-18, which uses the whole image as input, surpasses the three traditional machine learning-based methods (WBM, RBM, and PBM) with different input representations by significant margins in both tasks. It demonstrates that with the transfer learning from the massive medical dataset training \cite{Chen2019Med3D:}, Med3D can effectively extract useful high-level features from the entire sMRI image for the classification task. In addition, the deep learning-based methods learn high-level features from data in an incremental manner with a massive number of parameters and non-linear calculations, thus allowing better performance than traditional machine learning models.

\subsection{Results of Generalizability Evaluation }

\begin{table}[h]
\caption{Results of AD classification on the independent ADNI2 and 3 datasetss}\label{table:5}
\centering
\resizebox{0.45\textwidth}{!}{%
\begin{tabular}{lllll}
\hline
             & \multicolumn{4}{l}{AD   vs. NC classification}        \\ \hline
Model        & ACC         & SEN         & SPE         & AUC         \\
VBM          & 0.806±0.046 & 0.578±0.049 & 0.866±0.036 & 0.816±0.036 \\
RBM          & 0.789±0.102 & 0.522±0.103 & 0.865±0.102 & 0.795±0.076 \\
PBM          & 0.825±0.088 & 0.775±0.119 & 0.866±0.052 & 0.884±0.023 \\
Med3D-18     & 0.874±0.147 & 0.795±0.141 & 0.915±0.148 & 0.906±0.089 \\
HFCN         & 0.851±0.012 & 0.749±0.014 & 0.855±0.009 & 0.865±0.026 \\
DA-MIDL      & 0.868±0.034 & 0.772±0.092 & 0.893±0.101 & 0.901±0.097 \\
Our   method & 0.891±0.019 & 0.791±0.068 & 0.882±0.046 & 0.925±0.023 \\ \hline
\end{tabular}
}
\end{table}

{Table.~\ref{table:5}} shows the AD classification results of our method and the competing methods evaluated on the independent ADNI2 \&3 datasets. Our proposed sMRI-PatchNet generally outperforms the other five competing methods. the sMRI-PatchNet obtains the highest accuracy (0.891) in the AD vs. NC classification, outperforming VBM (0.806), RBM (0.789), PBM (0.825), Med3D-18 (0.874), HFCN (0.851) and DA-MIDL(0.868). These results indicate that the sMRI-PatchNet can provide robust performance across different datasets.

In general, the performance of a model is expected to decrease when evaluating on the independent dataset. The accuracy and AUC of our proposed model slightly decrease by 2\% and 4\%, respectively. These results indicate the good generalization capability of our method for AD diagnosis. The accuracy of machine learning-based methods, such as VBM, RBM, and PBM, only drops by around 1\%. This may be due to the fact that the ML-based methods have fewer parameters, allowing the model to avoid overfitting.

\subsection{The Impact of Discriminative Patch Location Selection on Model Performance}

\par
\begin{figure}[h]
    \centering
    \includegraphics[width=0.45\textwidth]{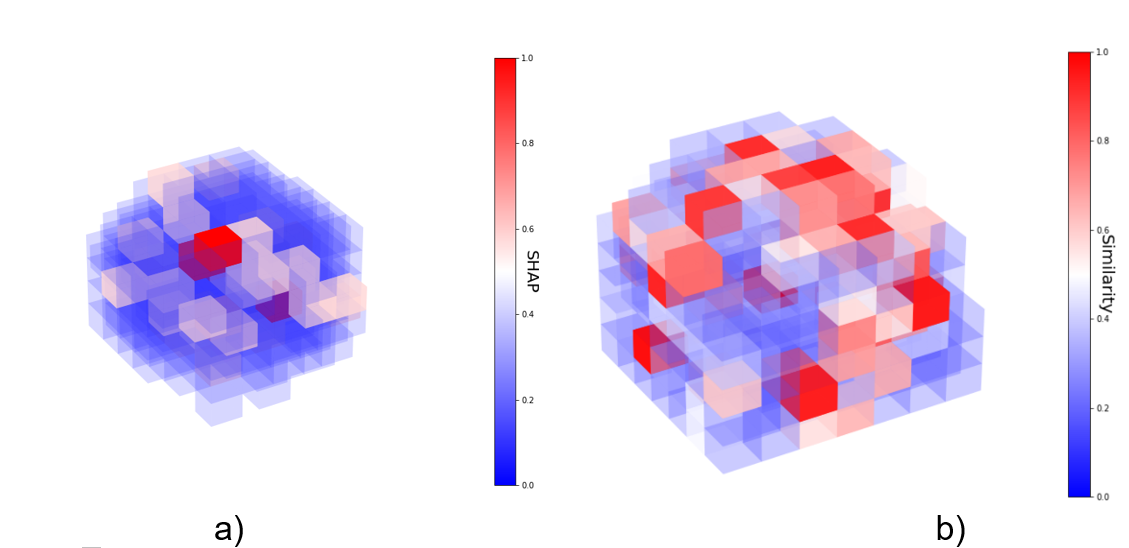}
    \caption{The discriminative patch location distribution determined by: (a) the proposed explainable SHAP based method and (b) the statistic analysis method.}
    \label{FIG:5}
\end{figure}

 {Fig.~\ref{FIG:5}} shows the distributions of discriminative patches selected by the statistical analysis method and the proposed explainable SHAP-based method. The discriminative patch locations determined by the SHAP-based method focus more on the central part of the sMRI image, while the results from the statistical analysis method are discrete and distributed in various regions. 

\begin{figure}[h]
    \centering
    \includegraphics[width=0.45\textwidth]{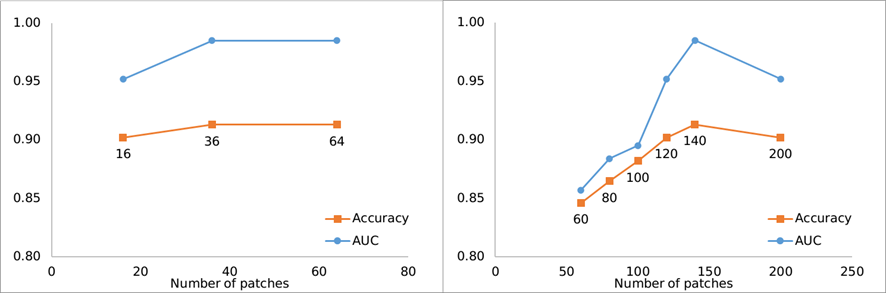}
    \caption{Results of Accuracy and AUC in AD classification obtained by: (a) our proposed explainable SHAP-based method and (b) the statistic analysis method with different selected numbers of input image patches.}
    \label{FIG:6}
\end{figure}

{Fig.~\ref{FIG:6}} shows the changes in the classification performance of our sMRI-PatchNet model with the increasing number of input image patches selected by the two methods, in terms of accuracy and AUC. It can be observed that PatchNet achieves satisfactory accuracy and AUC using the input patches selected by our proposed patch selection approach, even though the number of selected patches (n) is only 16. In contrast, the classification accuracy of PatchNet with the input patches selected by the statistic analysis method is only 0.846, after selecting a larger number (60) of input patches. Only under the circumstance of increasing the number of patches selected by the statistic method from 60 to 140, both the accuracy and AUC are significantly improved. This implies that the locations determined by the statistical methods are not necessarily correct, and a smaller number of patches are insufficient to yield satisfactory results. In our implementations, n = 36 is chosen to make a compromise between the computational complexity, the memory cost of training, and including a large enough number of potentially valuable locations.

\section{Discussions}

In this section, we first analyse the influence of the number and size of patch selection on the performance of the proposed model and its potential for clinical translation. Then, we compare our proposed method with previous studies on AD-related brain disease diagnosis. 

\subsection{Influence of Number and Size of Patches}

As a patch-based method, the size and number of patches are important parameters. We discuss the influence of the number and size of patches on the performance of the proposed model.  {Fig.~\ref{FIG:6}} a) shows the AD classification results achieved by the proposed model respectively with a range from 16 to 64. We can observe that both ACC and AUC are stable and better in the range of 16 to 64. The best performance is achieved when n increases to 36. This implies that small numbers of patches (e.g., n=16) may not include adequate patches related to AD classification. While, large numbers of patches (e.g., N=64) will increase the number of patches with useless information for AD classification.
\begin{figure}[h]
    \centering
    \includegraphics[width=0.45\textwidth]{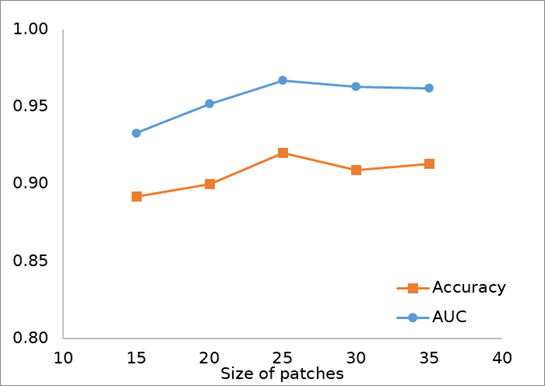}
    \caption{AD classification performance of the proposed model with the input patches of different sizes ( $15 \times 15 \times 15$ to $35 \times 35 \times 35$) on the ADNI test set.}
    \label{FIG:7}
\end{figure}
\par In this paper, we select a patch size of $25 \times 25 \times 25$, the same as used  \cite{Lian2020Hierarchical,Zhu2021Dual}. We evaluate the AD classification performance with different patch sizes.  {Fig.~\ref{FIG:7}} shows the AD classification performance with different patch sizes in a range of $15 \times 15 \times 15$ to $35 \times 35 \times 35$. The result shows the proposed model achieve a stable performance for all selected patches, and indicating that the model is not sensitive to the size of input patches within this range. The accuracy of the proposed model is greater than 0.9 for all selected patch sizes, except 15x15x15. This implies that a relatively large patch size is required in order to capture sufficient details of feature changes by brain atrophy.


\subsection{Discriminative Pathological Locations and the Potential of Clinical Translation}
\begin{figure*}[h]
    \centering
    \includegraphics[width=0.9\textwidth]{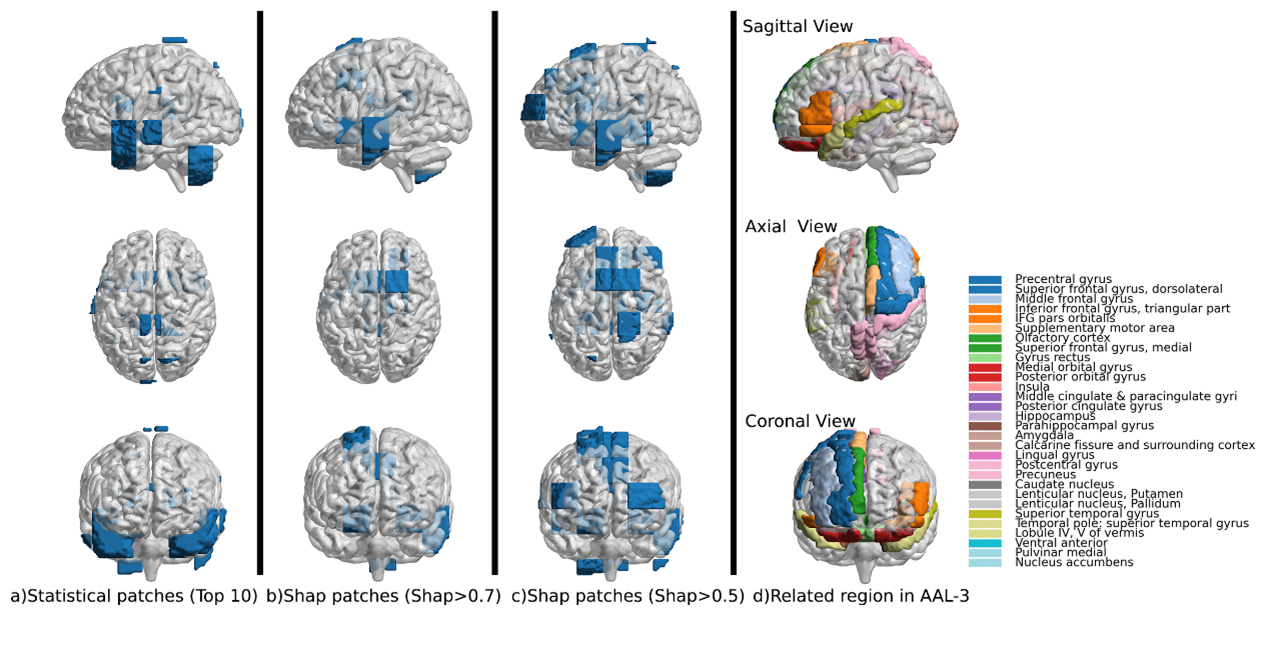}
    \caption{Discriminative AD-associated locations automatically identified by our proposed explainable SHAP-based method and the statistic analysis method. The first column shows the top 10 informative patch locations suggested by the statistical analysis. The second and third columns show the informative patch locations suggested by the proposed SHAP-based method with thresholds 0.7 and 0.5, respectively. The rightmost column shows the marked brain regions where the suggested patches are gathered by the proposed method.}
    \label{FIG:8}
\end{figure*}

\par In {Fig.~\ref{FIG:8}}, we visualize the locations of selected patches for AD diagnosis and their corresponding brain regions in the automated anatomical atlas (AAL 3V1) \cite{Rolls2020Automated}. Nearly half of the suggested discriminative locations by the statistic method are in the Posterior fossa, and the rest are in the cerebrum. 

\par However, the discriminative locations suggested by the proposed method cover 47 of 170 brain structures. {Table.~\ref{table:6}} lists 31 brain regions to which the patches suggested by our SHAP-based approach correspond in the brain atlas AAL 3V. They include Frontal Lobe, Temporal Lobe, Posterior Fossa, Insula and Cingulate Gyri, Occipital Lobe, Parietal Lobe, Central Structures, etc. These regions, such as the Precentral gyrus, Superior frontal gyrus, Middle frontal gyrus, Inferior frontal gyrus \cite{Yang2019Study}, Supplementary motor area \cite{Vidoni2012Evidence}, Olfactory cortex \cite{Murphy2019Olfactory}, Hippocampus, Parahippocampal gyrus, Amygdala \cite{Chen2016Differential}, Insula \cite{Foundas1997Atrophy}, Lingual gyrus \cite{Liu2017Decreased}, Precuneus \cite{Nelson2009Alzheimertype}, Caudate nucleus \cite{Persson2018Finding}, etc. are reported to be associated with AD. Specifically, the hippocampus is strongly linked to long-term memory. The impact of AD-related brain shrinkage on the hippocampus has been scientifically validated \cite{Planche2017Hippocampal}. The amygdala is considered to influence emotional functioning as well as learning and memory management \cite{Poulin2011Amygdala}. The thalamus is linked to cognition and information processing speed \cite{Aggleton2016Thalamic}, which are also relevant to AD. These pieces of evidence imply the feasibility of our proposed method for identifying AD-associated areas and can inform clinicians on AD diagnosis.

\begin{table}[h]
\caption{A list of brain regions that the suggested patches from our method cover in the automated anatomical atlas (AAL 3V1) \cite{Rolls2020Automated}.}\label{table:6}
\centering
\resizebox{0.45\textwidth}{!}{%
\begin{tabular}{llll}
\hline
Description                             & Code  & Description                              & Code    \\ \hline
\multicolumn{2}{l}{Frontal Lobe}                & \multicolumn{2}{l}{Insula and Cingulate Gyri}      \\
Precentral gyrus                        & 2     & Insula                                   & 33,34   \\
Superior frontal gyrus, dorsolateral    & 4     & Middle cingulate \& paracingulate gyri   & 38      \\
Middle frontal gyrus                    & 6     & Posterior cingulate gyrus                & 39,40   \\
Inferior frontal gyrus, triangular part & 9     & \multicolumn{2}{l}{Occipital Lobe}                 \\
IFG pars orbitalis                      & 11,12 & Calcarine fissure and surrounding cortex & 47      \\
Supplementary motor area                & 16    & Lingual gyrus                            & 52      \\
Olfactory cortex                        & 17,18 & \multicolumn{2}{l}{Parietal Lobe}                  \\
Superior frontal gyrus, medial          & 20    & Postcentral gyrus                        & 62      \\
Gyrus rectus                            & 23,24 & Precuneus                                & 71,72   \\
Medial orbital gyrus                    & 25,26 & \multicolumn{2}{l}{Central Structures}             \\
Posterior orbital gyrus                 & 29,30 & Caudate nucleus                          & 75,76   \\
\multicolumn{2}{l}{Temporal Lobe}               & Lenticular nucleus, Putamen              & 77,78   \\
Hippocampus                             & 41    & Lenticular nucleus, Pallidum             & 79,80   \\
Parahippocampal gyrus                   & 43,44 & Pallidum (PAL)                           & 81,82   \\
Amygdala                                & 45,46 & Ventral anterior                         & 126     \\
Superior temporal gyrus                 & 85    & Pulvinar medial                          & 145,146 \\
Temporal pole: superior temporal gyrus  & 87,88 & Nucleus accumbens                        & 158     \\
Posterior Fossa                         &       &                                          &         \\
Lobule IV, V of vermis                  & 115   &                                          &         \\ \hline
\end{tabular}
}
\end{table}

\subsection{Comparison with Previous Works}

For a broad comparison between our method and related studies on the performance of AD diagnosis, in Table 8 we list the results of several state-of-the-art models reported in the literature for AD classification and MCI conversion prediction tasks using structural MRI data from the ADNI database, including two voxel-level methods \cite{Cuingnet2011Automatic,Zhang2021Explainable}, two ROI-level methods \cite{Cao2017Nonlinearity-aware,Eskildsen2013Prediction} and four patch-level methods \cite{Lian2020Hierarchical,Liu2014Hierarchical,Tong2014Multiple,Zhu2021Dual}. The following observations can be noted in Table 8.
1) Our method has achieved a competitive performance in both AD-related classification tasks.
2) Compared with traditional machine learning-based methods such as SVM \cite{Cuingnet2011Automatic,Tong2014Multiple}, LDA \cite{Eskildsen2013Prediction}, and KNN \cite{Cao2017Nonlinearity-aware}, the deep learning-based methods have better performance, particularly for more difficult MCI conversion tasks. The possible reason is that deep learning methods have more parameters and can therefore deal with the spatial features and correlation of the 3D data better than machine learning methods. Compared to the other two deep learning patch-based methods, HFCN and DA-MIDL, the proposed method achieves better accuracy. As we mentioned in the Introduction section, the 3D convolution operation brings increased parameters and around six times the computational complexity (3×3 kernel size) than the 2D convolution operation. Table 6 shows the computational complexity and the number of parameters of the four deep learning-based methods. The Med3D with 10 layers, HFCN, DA-MIDL, and sMRI-PatchNet have a similar number of parameters (around 35 Million). The Med3D with 18 layers has the highest number of parameters (63.53 Million). However, in terms of computational complexity, the Med3D, HFCD and DA-MIDL are all use 3D convolution operation. Their computational complexities are 169.55 GMac and 240.73 GMac and 220.63 GMac, respectively. Our proposed sMRI-PatchNet uses 2D convolution and has the minimal computational complexity (2.21GMac). 
(3) Unlike ROI-based methods relying on  empirically predetermined ROIs, the proposed sMRI-PatchNet automatically extracts important areas from multiple patches distributed in the whole brain. This is much more difficult. However, our method still obtains good performance, implying the effectiveness of our model for identifying the location of pathology.

\begin{table*}[h]
\caption{Referential comparison on sMRI-based studies for ad classification and mci conversion prediction}\label{table:7}
\centering
\resizebox{0.9\textwidth}{!}{%
\begin{tabular}{lcllrrrrrr}
\hline
References               & \multicolumn{1}{l}{Feature}  & Method                      & Subject                     & \multicolumn{3}{c}{AD vs. NC   classification} & \multicolumn{3}{c}{pMCI vs.   sMCI classification} \\ \hline
\cite{Cuingnet2011Automatic}  & \multirow{2}{*}{Voxel-based} & SVM                         & 137AD+134sMCI+76pMCI+162NC  & 0.89           & 0.81          & 0.95          & 0.7              & 0.57           & 0.78           \\
\cite{Zhang2021Explainable}  &                              & 3D-CNN                      & 353AD+232sMCI+172pMCI+591NC & 0.91           & 0.91          & 0.92          & 0.82             & 0.81           & 0.81           \\
\cite{Eskildsen2013Prediction} & \multirow{2}{*}{ROIs-based}  & LDA                         & 194AD+234sMCI+161pMCI+226NC & 0.87           & 0.9           & 0.92          & 0.773            & 0.69           & 0.79           \\
\cite{Cao2017Nonlinearity-aware}      &                              & KNN                         & 192AD+229sMCI+168pMCI+229NC & 0.89           & 0.86          & 0.9           & 0.7              & 0.68           & 0.71           \\
\cite{Tong2014Multiple}      & \multirow{5}{*}{Patch-based} & SVM                         & 198AD+238sMCI+167pMCI+231NC & 0.9            & 0.86          & 0.93          & 0.72             & 0.79           & 0.74           \\
\cite{Liu2014Hierarchical}    &                              & Landmark detection + 3D CNN & 199AD+226sMCI+167pMCI+229NC & 0.91           & 0.88          & 0.94          & 0.77             & 0.42           & 0.82           \\
\cite{Lian2020Hierarchical}      &                              & Hierarchical-CNN            & 358AD+465sMCI+205pMCI+429NC & 0.88           & 0.89          & 0.88          & 0.81             & 0.81           & 0.8            \\
\cite{Zhu2021Dual}       &                              & Attention+MIL+CNN           & 398AD+232sMCI+172pMCI+400NC & 0.9            & 0.89          & 0.9           & 0.81             & 0.77           & 0.83           \\
Proposed Method          &                              & Custom CNN                  & 353AD+232sMCI+172pMCI+591NC & 0.92           & 0.92          & 0.92          & 0.82             & 0.82           & 0.86           \\ \hline
\end{tabular}
}
\end{table*}

\begin{table}[h]
\caption{The Computational complexity and parameters of proposed and Med3D methods}\label{table:8}
\centering
\resizebox{0.45\textwidth}{!}{%
\begin{tabular}{llll}
\hline
Model name        & Computational   complexity (Gflops) & Number   of parameters & Accuracy \\ \hline
Med3D   -10 Layer & 169.55   GMac                       & 35.47   Million        & 0.873    \\
Med3D   -18 Layer & 253.69   GMac                       & 63.53   Million        & 0.909    \\
HFCN              & 240.73   GMac                       & 37.86   Million        & 0.882    \\
DA-MIDL           & 220.63   GMac                       & 36.54   Million        & 0.904    \\
Our   Method      & 2.21   GMac                         & 34.53   Million        & 0.920    \\ \hline
\end{tabular}
}
\end{table}

\section{Conclusion}

This study has proposed a patch-based convolutional network with explainable patch location suggestions for Alzheimer’s Disease Diagnosis. First, we propose a fast and efficient explainable method for patch location suggestions through computing the SHapley Additive exPlanations (SHAP) contribution to a transfer learning model for AD diagnosis on massive medical data. A fast recursive partition perturbation method is introduced to effectively perturb the data to provide a fast estimation for the SHAP value of each patch. It has significantly reduced the number of patches required for achieving a good classification performance with 36 patches only, in contrast to 140 patches used in the existing statistical-based methods. Consequently, it dramatically reduces the computational complexity of the model, enabling efficient 3D data processing and analysis. Then, a novel patch-based convolutional network (sMRI-PatchNet) is designed to extract deep features of the discriminative patches and applied to AD classification and its transitional state moderate cognitive impairment (MCI) conversion prediction. The visualization results of brain regions covered by selected patches show that the proposed method can effectively identify discriminative pathological locations. These new biomarkers can help clinicians in clinical diagnosis. 
The classification performance and generalisability of our proposed method have been evaluated on two independent datasets and also compared with the five state-of-the-arts methods. The results show that the proposed model  outperforms the existing methods and has good generalizability in all cases. Moreover, it dramatically reduces the computational complexity and computational costs, compared to traditional deep learning methods. Future research will apply this method to more medical data and related disease diagnoses.

\appendices

\section*{Acknowledgment}

This work is supported by the Royal Society - Academy of Medical Sciences Newton Advanced Fellowship (NAF$\backslash$R1$\backslash$180371). L.H. Han was funded by the UK Engineering and Physical Science Research Council (EP/W007762/1) and the Small Business Research Initiative (Innovate UK, SBRI Funding Competitions: Heart Failure, Multi-morbidity and Hip Fracture).

\bibliographystyle{IEEEtran}
\bibliography{tem}

\end{document}